\pdfoutput=1
\RequirePackage{lineno}
\newif\ifpreprint%
\preprintfalse%
\ifpreprint%
	\documentclass[preprint,english,aps,prb,floatfix,amssymb,superscriptaddress,a4paper]{revtex4-1}
\else%
	\documentclass[twocolumn,english,aps,prb,floatfix,amssymb,superscriptaddress,a4paper]{revtex4-1}
\fi%
\usepackage{amsmath}
\usepackage{siunitx}
\usepackage{dsfont}
\usepackage{color}
\usepackage{amssymb}
\usepackage{graphicx}
\usepackage{chemformula}
\usepackage{braket}
\usepackage{units}
\usepackage{amssymb}
\usepackage{mathtools}
\usepackage{physics}
\usepackage{braket}
\usepackage{bibunits}

\usepackage{hyperref}
\hypersetup{
  colorlinks = true,
  hypertexnames=true
}
\renewcommand{\theta}{\vartheta}
\renewcommand{\phi}{\varphi}

\newcommand{\RN}[1]{%
  \textup{\uppercase\expandafter{\romannumeral#1}}%
}

\usepackage{times}

\newcommand{\beq}{\begin{equation}}
\newcommand{\eneq}{\end{equation}}
\newcommand{\bs}[1]{\boldsymbol{#1}}

\def\be{\begin{equation}}
\def\ee{\end{equation}}
\def\ba{\begin{eqnarray}}
\def\ea{\end{eqnarray}}

\def\R{{\rm Re}}
\def\Z{\mathbb{Z}}
\def\C{\mathbb{C}}

\def\s{\sigma}

\def\beq{\begin{equation}}
\def\eeq{\end{equation}}
\def\barray{\begin{eqnarray}}
\def\earray{\end{eqnarray}}

\font\upright=cmu10 scaled\magstep1
\def\stroke{\vrule height8pt width0.4pt depth-0.1pt}

\def\Zmath{\mathbb{Z}}
\def\Qmath{\vcenter{\hbox{\upright\rlap{\rlap{Q}\kern
                   3.8pt\stroke}\phantom{Q}}}}
\def\Nmath{\vcenter{\hbox{\upright\rlap{I}\kern 1.7pt N}}}
\def\Cmath{\vcenter{\hbox{\upright\rlap{\rlap{C}\kern
                   3.8pt\stroke}\phantom{C}}}}
\def\Rmath{\vcenter{\hbox{\upright\rlap{I}\kern 1.7pt R}}}
\def\Z{\ifmmode\Zmath\else$\Zmath$\fi}
\def\Q{\ifmmode\Qmath\else$\Qmath$\fi}
\def\N{\ifmmode\Nmath\else$\Nmath$\fi}
\def\C{\ifmmode\Cmath\else$\Cmath$\fi}
\def\R{\ifmmode\Rmath\else$\Rmath$\fi}

\input{epsf}

\newcounter{defcounter}
\begin{document}
\ifpreprint%
	\linenumbers%
\fi%

\title{Non-Abelian chiral spin liquid on a simple non-Archimedean lattice}

\author{
V.~Peri}
\address{
 Institute for Theoretical Physics, ETH Zurich, 8093 Zurich, Switzerland
}

\author{
S.~Ok}
\address{
 Department of Physics, University of Zurich, Winterthurerstrasse 190, 8057 Zurich, Switzerland
}

\author{
S.~S.~Tsirkin}
\address{
 Department of Physics, University of Zurich, Winterthurerstrasse 190, 8057 Zurich, Switzerland
}

\author{
T.~Neupert}
\address{
 Department of Physics, University of Zurich, Winterthurerstrasse 190, 8057 Zurich, Switzerland
}

\author{
G.~Baskaran}
\address{
Institute of Mathematical Sciences, Chennai 600 113, India
}
\address{
Perimeter Institute for Theoretical Physics, Waterloo, Ontario N2L 2Y5, Canada
}

\author{M.~Greiter}
\address{
 Institute for Theoretical Physics and Astrophysics, University of W\"urzburg, Am Hubland, D-97074 W\"urzburg, Germany
}

\author{R.~Moessner}
\address{Max-Planck-Institut f\"ur Physik komplexer Systeme,
  N\"othnitzer Str. 38, 01187 Dresden, Germany}

\author{R.~Thomale}
%\email{rthomale@physik.uni-wuerzburg.de}
\address{
 Institute for Theoretical Physics and Astrophysics, University of W\"urzburg, Am Hubland, D-97074 W\"urzburg, Germany
}

\let\oldaddcontentsline\addcontentsline
\renewcommand{\addcontentsline}[3]{}
\begin{bibunit}[prsty]
\begin{abstract}
We extend the
scope of Kitaev spin liquids to non-Archimedean lattices. For the pentaheptite lattice, which results from the proliferation of Stone-Wales defects on the honeycomb lattice, we
find an exactly solvable non-Abelian chiral spin liquid with spontaneous time reversal symmetry
breaking due to lattice loops of odd length. 
Our findings call for potential extensions of exact results for Kitaev
models which are based on reflection positivity, which is not
fulfilled by the pentaheptite lattice.  We further elaborate on
potential realizations of our chiral spin liquid proposal in strained $\alpha$-RuCl$_3$.
\end{abstract}

\date{\today}

\maketitle

Since the first proposal~\cite{doi:10.1080/14786439808206568}, quantum spin liquids have remained
an as fascinating as elusive direction of contemporary condensed
matter research on frustrated magnetism and topologically ordered
many-body states. In theory, different approaches have been
developed, many of which were
inspired by cuprate superconductors~\cite{ANDERSON1196} or the
fractional quantum Hall effect (FQHE)~\cite{PhysRevLett.59.2095}, but these were
limited due to the relative paucity of exactly solvable models~\cite{Kitaev:2003,Moessner:2001}. 
A fundamental breakthrough was reached by Kitaev in proposing a
microscopic Hamiltonian for quantum spin liquids with an emergent massive Ising gauge theory~\cite{Kitaev:2006}. Instead of just realizing
a desired spin liquid ground state wave function as an exact
eigenstate of a microscopic Hamiltonian, the powerful exact solution of the
Kitaev spin liquid allows for
the explicit analysis of anyonic excitations. Its solution is most elegantly
accomplished by a Majorana representation, where the eigenspectrum simplifies to a
free single-Majorana band structure. 

The Kitaev models realize both
Abelian and non-Abelian anyons~\cite{Kitaev:2006}, spontaneous time reversal symmetry
breaking chiral spin liquids~\cite{Yao:2007,Fu:2019}, a generalization to
$\mathbb{Z}_k$ gauge theory~\cite{PhysRevLett.114.026401}, and an extension to three-dimensional spin liquids
with anyon metallicity~\cite{Hermanns:2015,OBrien:2016,Miao:2018}. While the non-Abelian anyons in
the Kitaev model are of Ising type, alternative microscopic approaches to
non-Abelian spin liquids have found realizations of SU(2)$_k$ anyons
in
chiral~\cite{Greiter:2009,Greiter:2014,Meng:2015,PhysRevB.95.140406,PhysRevB.98.184409}
and non-chiral~\cite{PhysRevB.84.140404} spin liquids. 
The concept of spinon Fermi surface has been previously
developed in the context of Gutzwiller projections on fermionic mean
field states~\cite{PhysRevB.72.045105}. The exact solvability of the Kitaev models, however,
renders all these features accessible to an unprecedented degree, and
as such promises a more concise connection to observable
quantities~\cite{PhysRevLett.112.207203} and candidate materials~\cite{PhysRevLett.102.017205,PhysRevLett.108.127203,PhysRevB.90.041112,Hermanns:2018,Trebst:2017}.

\begin{figure}[t]
\includegraphics{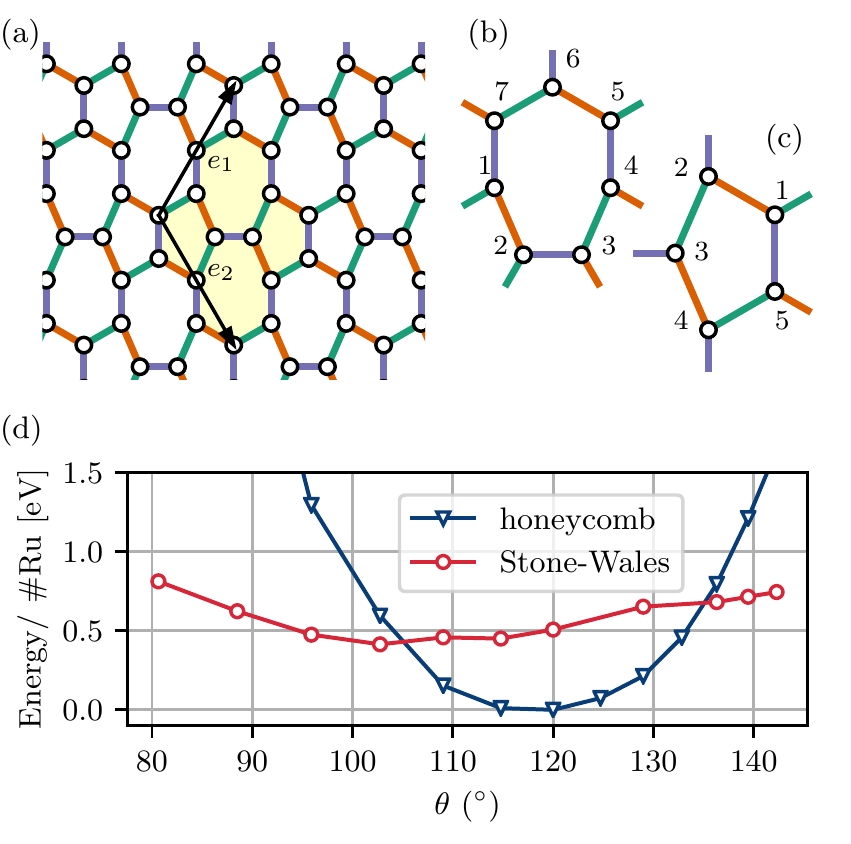}
\caption{(Color online) 
(a) pentaheptite lattice with a unit cell highlighted in light yellow
and the lattice vectors $\bs{e}_1$ and $\bs{e}_2$. Bond colors
highlight the type of spin-spin coupling across a bond $\sigma^{\alpha}_{j}\sigma^{\alpha}_{k}$, $\alpha=x,y,z$ (violet for $z$, orange for
$y$ and green for $x$). 
(b),(c) site labels for the definition of plaquette operators in~\eqref{eq: local conserved}. (d) Energy per \ch{Ru} atoms for $\alpha$-\ch{RuCl3} under strain with respect to the energy of the honeycomb configuration at $\theta=\ang{120}$. In red for a honeycomb lattice, in blue for a Stone-Wales defect. The strain changes the angle $\theta$ between the lattice vectors and the Stone-Wales structure is preferred at $\theta<\ang{105}$ and $\theta>\ang{135}$.
} 
\label{fig: lattice}
\end{figure}

In this work, we extend the Kitaev paradigm to non-Archimedean
lattices. Lattices can be classified by the
symmetry of sites and bonds. Archimedean lattices are formed by regular polygons where
each lattice vertex is surrounded by the same sequence of
polygons. This implies the equivalence of all sites, but not
necessarily of all bonds. Conversely, for lattices of the type including the Lieb lattice~\cite{Lieb:1989}, the symmetry-equivalence of all bonds does not imply the equivalence of sites.  In
a non-Archimedean lattice, neither all sites nor all bonds are
equivalent. As a prototypical example to which
we particularize in the following, pentaheptite (Fig.~\ref{fig: lattice}~(a))  exhibits irregular pentagons and
heptagons as well as two
types of vertices with $(5^1,7^2)$ and $(5^2,7^1)$ configuration,
respectively. In this notation $a^{m}$, the lattice is characterized by a list of the number of edges $a$ and the multiplicity $m$ of polygons that surround each inequivalent vertex. Pentaheptite~\cite{Crespi:1996,Deza:2000} can be thought of as originating from the
honeycomb $(6^3)$ lattice by the proliferation of
Stone-Wales defects. There, a pair of honeycomb bonded vertices change
their connectivity as they rotate by 90 degrees with respect to the
midpoint of their bond. The Stone-Wales defect proliferation transforms four contingent hexagons into two
heptagons and two pentagons. 
Pentaheptite has three-colorable bonds as the honeycomb lattice, and thus lends itself to an exact
solution of the Kitaev model, albeit not being three-colorable by faces.

\textit{Strain engineering of pentaheptite lattice in  $\alpha$-\ch{RuCl3} ---}
We perform first principle calculations of the candidate Kitaev honeycomb material $\alpha$-\ch{RuCl3} under uniaxial strain (see Supplemental Material [\onlinecite{supp}] and Ref.[\onlinecite{VASP1,VASP2,PAW1,PAW2,PBE-1996}] therein for more details). We find that under sufficiently strong tensile or compressive strain a configuration where the \ch{Ru} atoms arrange themselves in a pentaheptite lattice becomes favorable (see Fig.~\ref{fig: lattice}~(d)). This result motivates our choice to extend the exactly solvable Kitaev model to non-Archimedean lattices. Moderate strain in $\alpha$-\ch{RuCl3} has been shown to enhance magnetic Kitaev interactions~\cite{Biswas:2019}. It remains an open question whether the geometric frustration introduced by stronger strain~\cite{supp} is detrimental to the directional Kitaev exchange.

%%%%%%%%%%%%%%%%%%%%%%%%%%%%%%%%%%%%%%%%%%%%%%%%%%%%

\textit{Kitaev pentaheptite model ---}
We consider a  spin-1/2 degree of freedom on each pentaheptite
site. The unit cell shown in Fig.~\ref{fig: lattice}~(a) contains
eight sites and is spanned by the vectors $\bs{e}_1=(\sqrt{3}a,3a)$
and $\bs{e}_2=(\sqrt{3}a,-3a)$. We set the nearest-neighbor distance $a$ of
the underlying honeycomb lattice to unity. The Kitaev Hamiltonian reads
\be
H=
-J_x\sum_{x\text{-type}}\sigma_j^x\sigma_k^x  
- J_y\sum_{y\text{-type}}\sigma_j^y\sigma_k^y 
- J_z\sum_{z\text{-type}}\sigma_j^z\sigma_k^z\;,
\label{eq: model}
\ee
where $\sigma^{x,y,z}_j$ denotes the Pauli matrices acting on site $j$, the sums run over distinct sets of bonds connecting nearest
neighbor sites $j$ and $k$, and $J_{x,y,z}\in \mathbb{R}$. Which bonds contribute to each sum is
shown in Fig.~\ref{fig: lattice}~(a).
Each heptagon and pentagon of the lattice is associated with a conserved quantity of~\eqref{eq: model} given by
\begin{align}
  \begin{split}
W_{\text{pen}}&=K_{12}K_{23}K_{34}K_{45}K_{51}\;,\\%\s^x_1 \s^y_2 \s^z_3 \s^z_4 \s^z_5\;,\quad
W_{\text{hep}}&=K_{12}K_{23}K_{34}K_{45}K_{56}K_{67}K_{71} \;,%\s^x_1 \s^y_2 \s^z_3 \s^x_4 \s^y_5  \s^y_6 \s^x_7
\label{eq: local conserved}
\end{split}
\end{align}
where $K_{ij}=\s_i^\alpha\s_j^\alpha$ for sites $i$ and $j$ connected by a bond of type $\alpha=x,y,z$.
The spin operators act on the sites around each pentagon and heptagon
according to the site labels in Fig.~\ref{fig:
  lattice}~(b) and~(c), respectively. The conserved quantities for the
pentagons and heptagons that relate to those shown
in Fig.~\ref{fig: lattice}~(b) and~(c) via mirror reflection are defined analogously.
All $W_{\text{pen}}$ and $W_{\text{hep}}$ commute
with each other and the Hamiltonian~\eqref{eq: model},  which can thus
be diagonalized in each eigenspace of these operators (``flux
sector'') separately.
 
Importantly, in contrast to the Kitaev honeycomb case,
time-reversal $T$ commutes with the plaquette operators $W_l$ but flips their eigenvalues. Applying $T$ to the equation $W_l\left|\psi\right\rangle=w_l\left|\psi\right\rangle$, one gets $W_lT\left|\psi\right\rangle=w_l^*T\left|\psi\right\rangle$. The reason is that the elementary loops in the
pentaheptite lattice are of odd length and have imaginary eigenvalues
$\pm\mathrm{i}$. This implies 
spontaneous time reversal symmetry breaking. In particular, one needs to specify the direction followed around the plaquette and in definition~\eqref{eq: local conserved} we choose a counterclockwise convention. A similar situation is found
on the decorated honeycomb lattice $(3,12^2)$ of the Kitaev-Yao-Kivelson (KYK) model~\cite{Yao:2007}.  In our case, however, \emph{all} elementary loops are of odd length.
%The plaquette operators defined in~\eqref{eq: local conserved} ensure that the composition law for fluxes holds~\cite{PhysRevB.90.134404}: $W_{1+2}=W_1W_2$. 

With the conserved plaquette quantities identified, one can map the
system to noninteracting Majorana fermions in each flux sector
by following Kitaev's procedure~\cite{Kitaev:2006}: we replace each spin (site $j$) by four Majorana fermions $c_j$, $b_j^\alpha$, $\alpha=x,y,z$, and restrict the Hilbert space to that of even fermion parity on each site\cite{Pedrocchi:2011}. The resulting Hamiltonian takes the form
\begin{equation}
H=\frac{\mathrm{i}}{4}\sum_{\langle j k\rangle}A_{jk} c_j c_k,
\label{HMajo}
\end{equation}
where the sum runs over nearest-neighbor sites and
$A_{jk}=  2J_{\alpha_{jk}}u_{jk}$ if sites $j$ and $k$ are connected by an $\alpha$ link, $\alpha_{j,k}\in\{x,y,z\}$.
The Majorana bilinears $u_{jk}=\mathrm{i} b_j^{\alpha_{jk}}b^{\alpha_{jk}}_k$
commute with each other and the Hamiltonian. As Hermitian operators that square to 1, we can replace them by their eigenvalues $\pm1$. Their eigenvalues are related to those of $W_{\text{pen}}$ via
\begin{equation}
W_{\text{pen}}=(-\mathrm{i})^5\prod_{\langle j k\rangle\in\text{pen}} u_{jk},
\label{eq: pen op in Majorana}
\end{equation}
where the product runs over all bonds that form a given pentagon. The
analogous equation holds for $W_{\text{hep}}$ with the prefactor $(-\mathrm{i})^7$. 
Thus, while the
eigenvalue of $u_{jk}$ on a given bond is gauge variant, the product
of eigenvalues around a closed loop is a gauge invariant
$\mathbb{Z}_2$ flux. Note that the order in the product of Eq.~\eqref{eq: pen
  op in Majorana} requires to fix a positive direction for the bond operators $u_{jk}$~\cite{supp}. According to our convention, the configuration with all the $u_{ij}$ with positive eigenvalues corresponds to heptagonal (pentagonal) plaquettes with eigenvalue $+\mathrm{i}$ ($-\mathrm{i}$).

\begin{figure}[t]
\includegraphics{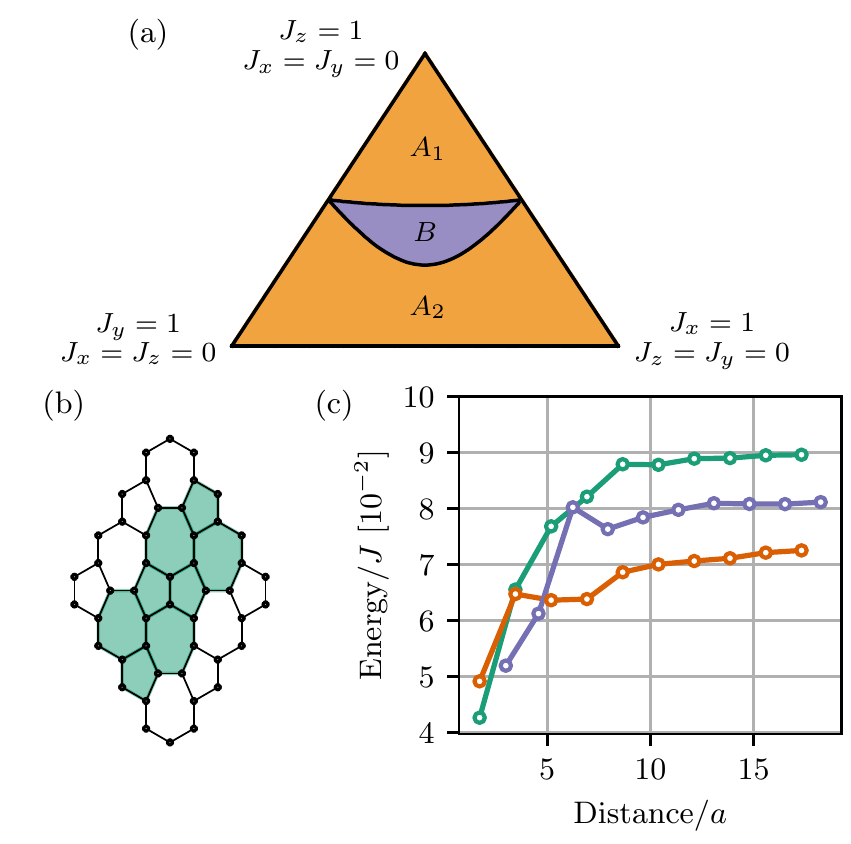}
\caption{
(a) Phase diagram of the vortex-free sector of Hamiltonian~\eqref{HMajo}. Phases $A_1$ and $A_2$ are topologically equivalent and realize the Abelian $\mathbb{Z}_2$ topological order. Phase $B$ realizes the non-Abelian Ising topological order.
(b) Flux configuration of lowest excited state for a system on the torus $L\bs{e}_1\times L\bs{e}_2$ with $L=2$. Colored plaquettes have flux $\pi$.
(c) Energy cost per vortex for a pair of vortices as a function of their separation on a torus with $L=10$ and $J_x=J_y=J_z=J$. For the green curve, vortices are in the pentagons. Vortices in heptagons are shown in orange. For the violet curve, one vortex is in a heptagon and the other in a pentagon. 
} 
\label{fig: vortex sector}
\end{figure}

\textit{Identifying the ground state flux sector ---}
For each flux sector, the ground state energy of Eq.~\eqref{HMajo} can
be determined.
A powerful
result by Lieb~\cite{Lieb:1994}, based on reflection positivity,
assures that if a Kitaev-type spin model possesses reflection symmetry
such that the plane of reflection does not contain any lattice site,
a ground state is always in the flux free sector. The Kitaev model on
the pentaheptite lattice is particular in that it does not have such a mirror symmetry.  

From our flux sector analysis, we conjecture that even for Eq.~\eqref{eq: model}, the
ground state is in the flux free sector, i.e., the sector where all $u_{ij}$ have positive eigenvalues according to the chosen convention. Numerical evidence along this
line has been provided for other systems lacking reflection
positivity, while a rigorous result is missing
\cite{Chesi:2013}. Fixing $J_x=J_y=J_z=1$ without loss of
generality, we find the following: (i)
  For a cluster of $2\times 2$ unit cells, the energy computed for all
  vortex configurations~\cite{supp} singles out the
  vortex-free sector with energy $-3.1044$ 
  per unit cell and an excitation gap of
  $0.0106$. 
 The first excited sector is a cluster of $\pi$ vortices in
  half of the plaquettes as displayed in Fig.~\ref{fig: vortex
    sector}~(b). This first excited state is particularly affected by finite-size effects, while we conjecture that the first
	excited state in larger samples is realized by a single pair of vortices close to each other.~\cite{supp} (ii) Upon increasing system size, the energy of
the vortex-free configuration extrapolates to $-3.0971$ in the
thermodynamic limit. (iii) The energy cost of nucleating a
pair of vortices tends to a nonzero constant with increasing separation, indicating
that it is not energetically favorable to nucleate isolated vortices
[Fig.~\ref{fig: vortex sector}~(c)]. Based on this numerical evidence,
we consider the vortex-free sector for our subsequent analysis of the
pentaheptite Kitaev spectrum. 

%\begin{figure}[t]
%\includegraphics{./figures/Figure-3.pdf}
%\caption{
%Boundary spectrum of the flux free sector of  model~\eqref{eq:
%  model} for a ribbon with open boundary
%conditions along $\bs{e}_1$ and periodic along $\bs{e}_2$, with $J_x = J_y = J_z
%= J$. The ribbon is composed of 50 unit cells along $\bs{e}_1$. The Chern
%numbers of the three gaps are  $C=-1,+1,-1$ from bottom to top. Modes
%localized at one edge are plotted in green and those localized
%at the other edge in orange. The inset shows the edge terminations of the ribbon with respective colors.
%} 
%\label{fig: boundary spectrum}
%\end{figure}

\textit{Phase diagram ---}
The state with all $W_{\text{hep}}=+\mathrm{i},\,W_{\text{pen}}=-\mathrm{i}$ and that with
$W_{\text{hep}}=-\mathrm{i},\,W_{\text{pen}}=+\mathrm{i}$ are degenerate and related by
time-reversal symmetry. Thus, in the vortex-free sector, the system
spontaneously breaks time-reversal symmetry. Without loss of
generality, we discuss the phases for $J_{x,y,z}>0$, as the sign of
the couplings is irrelevant: A change in the sign of $J_x$ or $J_y$
can be reabsorbed by changing the sign of an even number of $u_{jk}$ per
plaquette without adding vortices. At the same time, $J_z<0$ can be
mapped to a configuration with $J_z>0$ and an odd number of $u_{jk}$'s per
plaquette with flipped signs. This move adds a vortex in each plaquette, sending each
configuration to its time reversed partner, and does not affect its
energy~\cite{supp}. 

As shown in the ternary phase diagram Fig.~\ref{fig: vortex
    sector}~(a), we find three gapped phases, which are separated by first-order phase transitions~\cite{supp} at the
phase boundaries given by
\begin{equation}
J_z^2=J_x^2\pm\sqrt{2}
    J_x J_y +J_y^2,
\end{equation}
where the gap in the Majorna single particle spectrum closes.

Phases $A_1$ and $A_2$ are conveniently understood in a limit where
one of the couplings $J_x$, $J_y$, or $J_z$ is much larger than the
others. This is a good starting point for a perturbation theory in the Majorana fermion representation~\cite{PhysRevB.90.134404}. One finds~\cite{supp} that only non-contractible loops give a flux-dependent correction to the energy. In particular, a loop with $n$ weak bonds of strength $J$ gives a correction of order $J^n$. Moreover, loops of odd length do not give any shift in energy as their contribution is cancelled by their time reversal partners. Assume $J_z\gg J_x, J_y$ in $A_1$. The first non-trivial correction is given by the loops of length 10 involving adjacent pentagonal and heptagonal plaquettes. The Hamiltonian in sixth order perturbation theory reads
\be
\label{A1Effective}
H^{(6)}_{\text{eff}}=\text{const.}\,-\frac{7}{128}\left(\frac{J_x^4J_y^2}{|J_z|^5}+\frac{J_x^2J_y^4}{|J_z|^5}\right)W_{\text{pen}}W_{\text{hep}}\,.
\ee
In $A_2$, assume $J_y\gg J_x, J_z$ without loss of generality as the model is symmetric under $J_x\leftrightarrow J_y$. The first non trivial contribution arises in fourth order perturbation theory from the loops of length 8 involving two pentagons. This correction does not provide information on the energy cost of vortices in the heptagonal plaquettes. This enters in sixth order perturbation theory via the loops of length 10 enclosing a pentagon and a heptagon. The perturbative Hamiltonian up to sixth order for the phase $A_2$ is:
\be
\label{A2Effective}
H^{(6)}_{\text{eff}}=\text{const.}\,+\frac{5J_x^4}{16|J_y|^3}W_{\text{pen}^{}}W_{\text{pen}^{'}}-\frac{7J_x^2J_z^4}{128|J_y|^5}W_{\text{pen}}W_{\text{hep}}\,.
\ee
From~\eqref{A1Effective} and~\eqref{A2Effective}, we see that both in $A_1$ and in $A_2$ the vortex sector is gapped and the ground state is in the flux-free sector, i.e., $W_{\text{hep}}=+\mathrm{i}$ and $W_{\text{pen}}=-\mathrm{i}$.  
To study the vortex excitations, consider the phase $A_1$ in the limit $J_z\gg J_x, J_y$. A pair of vortices can be created in two heptagonal plaquettes or in two pentagonal ones. The energy of these vortices shows little dependence on their separation. 
On the other hand, a single pair of vortices in a heptagon and a pentagon changes the fermionic parity and it is thus unphysical. These vortices do not carry unpaired Majorana modes. Similar results hold for the $A_2$ phase.
These observations, together with the four-fold ground state degeneracy, reproduce the fusion rules and the topological degeneracy of $\mathbb{Z}_2$ topological order~\cite{Kitaev:2006} and support the claim that these phases realize the same topological order as the same limit of Kitaev models on Archimedean lattices. 

\begin{figure}[t]
\includegraphics{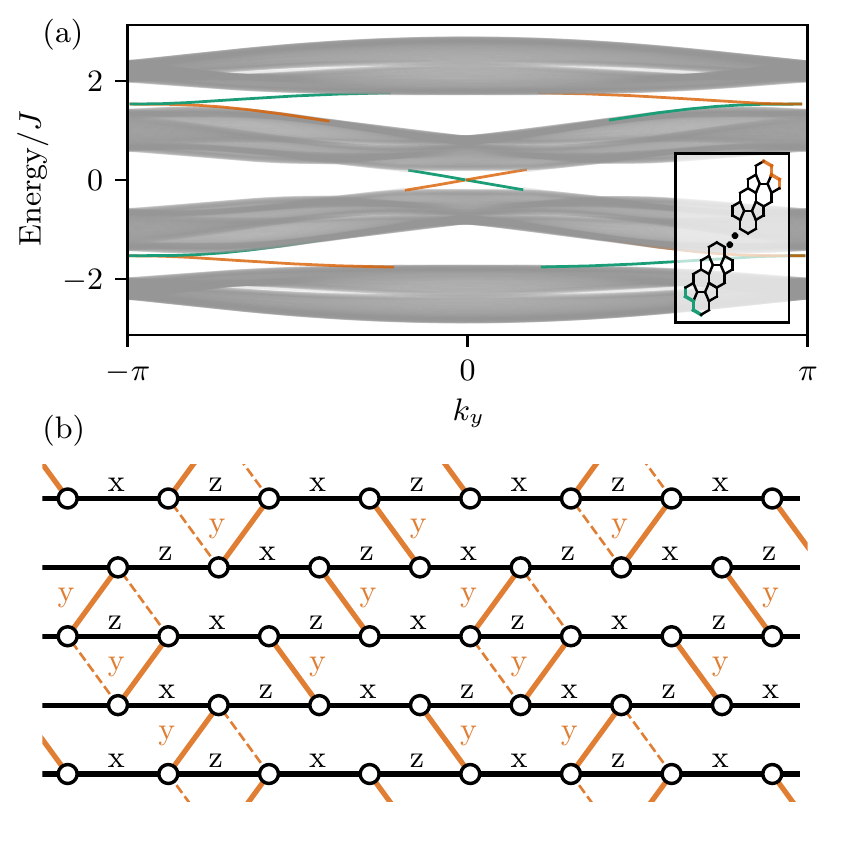}
\caption{
(a) Boundary spectrum for a ribbon with open boundary
conditions along $\bs{e}_1$ and periodic along $\bs{e}_2$, with $J_x = J_y = J_z
= J$. The ribbon is composed of 50 unit cells along $\bs{e}_1$. The Chern
numbers of the three gaps are  $C=-1,+1,-1$ from bottom to top. Modes
localized at one edge are plotted in green and those localized
at the other edge in orange. The inset shows the edge terminations of the ribbon with respective colors. (b) 
Non-Abelian chiral spin liquid (phase $B$) arising from weakly coupled spin chains. 
The Kitaev pentaheptite model can be deformed to an array of spin chains with alternating $\sigma^x$-$\sigma^x$ and $\sigma^z$-$\sigma^z$ coupling 
that are weakly coupled via $\sigma^y$-$\sigma^y$ interactions (orange lines). The dashed lines represent the $\sigma^y$-$\sigma^y$ interactions for the coupled wire construction of the honeycomb lattice. 
} 
\label{fig: boundary spectrum}
\end{figure}

%\begin{figure}[t]
%\includegraphics{./figures/Figure-4.pdf}
%\caption{
%Non-Abelian chiral spin liquid (phase $B$) arising from weakly coupled spin chains. 
%The Kitaev pentaheptite model can be deformed to an array of spin chains with alternating $\sigma^x$-$\sigma^x$ and $\sigma^z$-$\sigma^z$ coupling 
%that are weakly coupled via $\sigma^y$-$\sigma^y$ interactions (orange lines). The dashed lines represent the $\sigma^y$-$\sigma^y$ interactions for the coupled wire %construction of the honeycomb lattice. 
%} 
%\label{fig: coupled wires}
%\end{figure}

Phase $B$, which also contains the isotropic point $J_x=J_y=J_z$, is
the chiral non-Abelian spin liquid. Our numerical studies suggest that
both the vortex sector and the fermionic sector of this phase are
gapped (see Supplemental Material and Fig.~\ref{fig: vortex sector}~(c)). Hence, vortices have well defined statistics. This can be
entirely determined by the Chern number $C$ associated to the
Majorana spectrum according to the sixteen-fold way for Majorana
fermions in a $\mathbb{Z}_2$ background gauge
field~\cite{Kitaev:2006}. We find~\cite{Fukui:2005} that the spectral
gap at half filling has $|C|= 1$. An odd Chern number is linked to
non-Abelian statistics of the vortex excitation which carries an
unpaired Majorana zero mode (MZM). In the presence of well isolated
vortices, these MZMs can be resolved already via exact diagonalization. A pair of MZMs $\{\gamma_i\,,\gamma_j\}$ can be used to construct a non-local fermionic degree of freedom $a=1/2(\gamma_i+\mathrm{i}\gamma_j)$ with an associated two dimensional Fock space, such that a system of isolated $2n$ vortices possess a topological degeneracy $2^n$. Taking into account the non-contractible loops on the torus and imposing fermionic parity conservation, the topological degeneracy for $2n$ vortices in the $B$ phase is $2^{n+1}$.  
%The topological spin of the vortex excitation is $\theta=\text{exp}(\pm \mathrm{i}\pi/8)$ and the associated quantum dimension $d=\sqrt{2}$. 
Therefore, the chiral non-Abelian spin liquid of the $B$ phase is linked to Ising field theory. It
is the same phase that can be induced by a magnetic field in Kitaev's
model on the honeycomb lattice, albeit in this case at the cost of
exact solvability~\cite{Kitaev:2006}. Here, it is accompanied by a
spontaneous breaking of time-reversal symmetry (by choosing all fluxes
to be $0$ or $\pi$), as it is the case for the KYK model 
%on the $(3,12^2)$ Archimedean lattice 
in Ref.~\onlinecite{Yao:2007}. Other possible realizations of this phase include the $\nu=5/2$ FQHE~\cite{Moore:1991} and 2D topological superconductors~\cite{Ivanov:2001}. 
The exact solubility of the model in the $B$ phase offers the opportunity to study
its chiral topological edge states for a geometry with open boundary
conditions, as presented in Fig.~\ref{fig: boundary spectrum}~(a). The
boundary theory of the Ising topological order is a single chiral
Majorana fermion mode, in accordance with $|C|=1$.

\textit{Coupled wire limit ---}
The limit $J_z=J_x\gg J_y$ is particularly interesting to study the gapped non-Abelian chiral spin liquid phase. 
In this limit, \eqref{eq: model} can be viewed as a
collection of critical one-dimensional Ising chains with
alternating $x$- and $z$-type terms (see Fig.~\ref{fig: boundary spectrum}~(b)) that are weakly coupled with $y$-type terms~\cite{Huang:2016,Huang:2017}. The same limit
can be considered for the original Kitaev honeycomb model, which leads
to a brick-wall lattice of weakly coupled chains. Thus, by merely
changing the geometry of how the chains are connected, one
can go from the Abelian Kitaev honeycomb model to a non-Abelian
Kitaev model. This represents one of the main advantages of the pentaheptite lattice over the KYK model. In fact, the latter cannot be obtained by a simple coupled wire construction, since in the non-Abelian limit $J'\ll J$ (see Ref.~\onlinecite{Yao:2007}), it consists of disconnected triangles. Recent ideas using Majorana-fermion based ``topological hardware''
offer a promising route toward realizing topologically ordered spin
models (the ``topological software'')\cite{Sagi:2018,
  Landau:2016,Plugge:2016}.

%one can turn
%the gapless phase of the Kitaev honeycomb model into the gapped non-Abelian chiral spin liquid.  

\textit{Extensions to 3D---}
Non-Archimedean lattices in three dimensions are abundant and go
beyond the classification studied in
Refs.~\onlinecite{OBrien:2016,Wells:1977}. 
%While we defer a detailed complete analysis to
%future work, 
It is interesting to notice how the pentaheptite lattice
has a natural 3D extension in a three-coordinated lattice with elementary
loops of odd length, e.g., the $(9,3)\text{a}$ lattice~\cite{OBrien:2016}. As such, it is amenable to an exact solution of
the Kitaev model and spontaneously breaks time reversal symmetry,
as the pentaheptite lattice in 2D. 
%The Majorana spectrum of this
%model has Weyl points in the zero flux sector in the absence of external perturbations that explicitly break time reversal symmetry. 

The $(9,3)\text{a}$ lattice has so far been understood mainly in terms
of stacked honeycomb layers via mid-bond sites. It can, however,
alternatively be obtained from the pentaheptite lattice by replacing the bonds shared by a pair of heptagons along $\bs{e}_1-\bs{e}_2$ with triangular spirals. 
The fact that the non-Archimedean 2D lattice studied here can
originate from a simpler Archimedean 3D lattice may pave the way to candidate materials for this model which have not been considered
previously, and stresses that non-Archimedean systems may generically
arise from the dimensional reduction of an Archimedean parent lattice.

\textit{Summary ---} We have generalized the Kitaev spin liquid paradigm to
non-Archimedean lattices, and find the Kitaev pentaheptite model to host an Ising-type
non-Abelian chiral spin liquid. Towards a possible realization of this
state of matter, we find that the Kitaev honeycomb material
$\alpha$-\ch{RuCl3} forms the pentaheptite lattice under uniaxial
strain. A future challenge will be to accomplish this experimentally, e.g., by substrate
engineering; and, from a complementary theoretical point of view, to provide a microscopic
derivation of the magnetic exchange interactions in the
modified pentaheptite structure. More broadly, while there has been some systematic work on frustrated 
magnetism on Archimedean lattices~\cite{Farnell:2014}, a comparable effort on both Lieb-type and 
non-Archimedean lattices is still lacking, and we hope that our work will motivate
such studies.

%%%%%%%%%%%%%%%%%%%%%%%%%%%%%%%%%%%%%%%%%%%%%%%%%%%%
\textit{Acknowledgments ---}
We thank Oleg Tchernyshyov for helpful comments and discussions.  VP acknowledges support from the European Research
Council under the Grant Agreement No. 771503 and NCCR MaNEP of the Swiss National
Science Foundation. SO was supported by the Swiss National Science
Foundation under grant 200021\_169061. TN and SST acknowledge support from
the European Unions Horizon 2020 research and innovation program
(ERC-StG-Neupert-757867-PARATOP). GB acknowledges the support of the
Wilhelm Wien institute as a distinguished Wilhelm Wien professor at the University
of W\"urzburg, where this work was initiated, DST-SERB (India) for SERB Distinguished Fellowship and research at Perimeter Institute, as a DVRC, is supported by the Government of Canada through Industry Canada and by the Province of Ontario through the Ministry of Research and Innovation. The work in W\"urzburg and Dresden is funded by the Deutsche Forschungsgemeinschaft (DFG, German Research
   Foundation) through Project-ID 258499086 - SFB 1170, Project-ID 247310070 - SFB 1143, and through the
   W\"urzburg-Dresden Cluster of Excellence on Complexity and Topology
   in Quantum Matter -- \textit{ct.qmat} Project-ID 39085490 - EXC
   2147.
%\putbib[biblioNew]

\end{bibunit}
\let\addcontentsline\oldaddcontentsline

\renewcommand{\thetable}{S\arabic{table}}
\newcommand{\ph}{\phantom\dagger}
\renewcommand{\thefigure}{S\arabic{figure}}
\renewcommand{\theequation}{S\arabic{equation}}
\onecolumngrid
\pagebreak
\setcounter{page}{1}
\thispagestyle{empty}
\begin{center}
	\textbf{\large Supplemental Material: Non-Abelian chiral spin liquid on a simple non-Archimedean lattice}\\[.2cm]
	
	  V.~Peri,$^{1}$ S.~Ok,$^{2}$ S.~S.~Tsirkin,$^{2}$ T.~Neupert,$^{2}$ G.~Baskaran,$^{3,4}$ M.~Greiter,$^{5}$ R.~Moessner,$^{6}$ and R.~Thomale$^{5}$ \\[.1cm]
	  {\itshape ${}^1$Institute for Theoretical Physics, ETH Zurich, 8093 Zurich, Switzerland\\
	  ${}^2$Department of Physics, University of Zurich, Winterthurerstrasse 190, 8057 Zurich, Switzerland\\
	   ${}^3$Institute of Mathematical Sciences, Chennai 600 113, India\\
	    ${}^4$Perimeter Institute for Theoretical Physics, Waterloo, Ontario N2L 2Y5, Canada\\
		 ${}^5$Institute for Theoretical Physics and Astrophysics, University of W\"urzburg, Am Hubland, D-97074 W\"urzburg, Germany\\
		 ${}^6$Max-Planck-Institut f\"ur Physik komplexer Systeme, N\"othnitzer Str. 38, 01187 Dresden, Germany\\}
	(Dated: \today)\\[1cm]
	\end{center}
	\appendix
	\renewcommand{\thesection}{\Roman{section}}
	\begin{bibunit}[prsty]
	\begin{center}
\textit{Physical subspace projection}
\end{center}
The states obtained in the
Majorana representation are projected to the physical subspace formed by $\ket{\psi}$s that satisfy $D_i\ket{\psi}=\ket{\psi}$ for each site index $i$, where $D_i=b_i^xb_i^yb_i^zc_i$. The projector on the physical subspace can be written as:
\begin{equation}
\mathcal{P}=\prod_{i=1}^{8N}\left( \frac{1+D_i}{2}\right)\;,
\end{equation}
where $N$ is the number of unit cells in the system. As highlighted in
Ref.~\onlinecite{Pedrocchi:2011}, unphysical states can likewise be
characterized by the positive eigenvalue of the operator
$\prod_{i=1}^{8N} D_i$, which is more convenient to compute. Restricting our attention to physical states, each configuration of
the system with periodic boundary conditions is entirely characterized
by the eigenvalues of the elementary plaquette operators and two additional
numbers $(\phi_x\,,\phi_y)$.  These are two global $\mathbb{Z}_2$
fluxes computed along a contour that crosses the whole sample along $x$ and $y$, respectively.

\begin{center}
\textit{DFT calculations of $\alpha$-\ch{RuCl3}}
\end{center}
The structural relaxation and total-energy calculations of $\alpha$-\ch{RuCl3} under uniaxial strain was performed  
within the framework of density functional theory (DFT) with the scalar-relativistic approximation. 
The calculation is done using  the {\tt VASP} code~\cite{VASP1,VASP2} employing the projector-augmented wave
method (PAW) \cite{PAW1,PAW2} and  the Perdew, Burke, and Ernzerhof generalized-gradient 
approximation (GGA-PBE)~\cite{PBE-1996} for the exchange-correlation functional. 
We start from the honeycomb lattice of $\alpha$-\ch{RuCl3} monolayer with equilibrium lattice parameter, 
and then change the angle $\theta$ between the unit cell vectors,
simultaneously adjusting the lattice parameter to keep the surface of the 2D unit cell fixed. 
For each value of $\theta$, all atoms are relaxed to their equilibrium positions inside the unit cell, yielding the minimal total energy. 
Further, we take a $2\times 2$ supercell to introduce one Stone-Wales (SW) defect and relax the structure for 
different values of the stretching angles $\theta$, in the same way as for the honeycomb lattice. 

\begin{figure}[!h]
\includegraphics{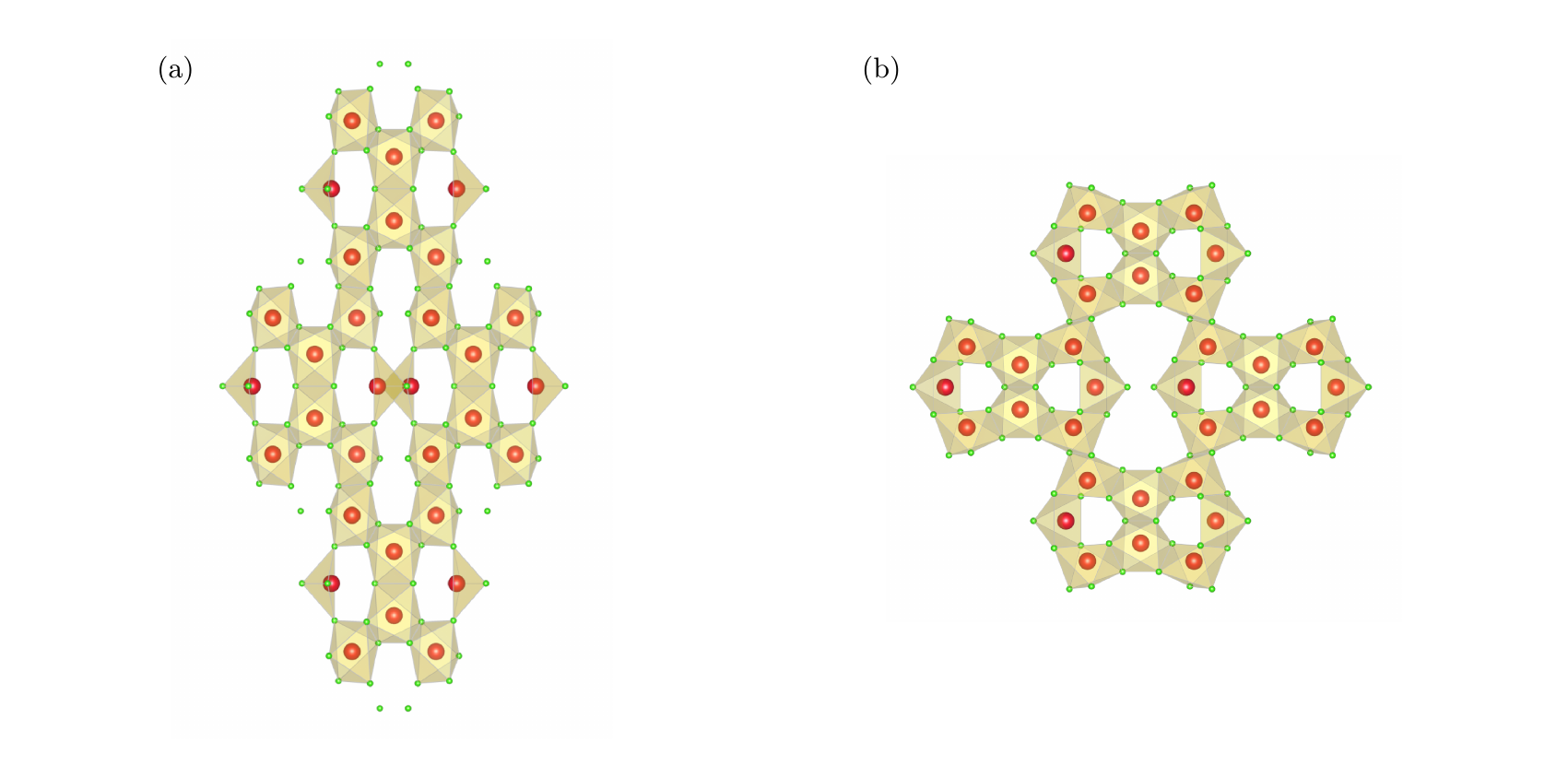}
\caption{
Crystal structure of the Kitaev material $\alpha$-\ch{RuCl3} under strain. (a) Angle between lattice vectors $\theta=\ang{136}$. (b) Angle between lattice vectors $\theta=\ang{96}$.
}  
\label{fig:DFT}
\end{figure}

The resulting total energies are shown in Fig.~1~(a) of the Main Text. 
One can see that at $\theta=\ang{120}$ the honeycomb lattice is energetically preferred over the Stone-Wales structure. 
We take the minimum energy of the honeycomb structure as a reference point for Fig.~1~(a) of the Main Text.
As we change the angle $\theta$ from $\ang{120}$, the  energy sharply increases. On the other hand, the energy of the SW structure 
has a weaker dependence on the angle, and thus becomes energetically preferred at $\theta<\ang{105}$ and $\theta>\ang{135}$.
Moreover, the equilibrium angle for this structure is around $\ang{102}$, which gives a hope for an experimental realization.
The atomic structures for $\theta=\ang{136}$ and $\theta=\ang{96}$ are shown in Fig.~\ref{fig:DFT}~(a) and~(b), respectively.

Under sufficient strain, the pentaheptite lattice studied in this work is favorable. Although the DFT calculations have been performed on the candidate Kitaev material $\alpha$-\ch{RuCl3}, it is not obvious whether Kitaev-like exchanges continue to be relevant. The number of \ch{Ru} atoms and octahedral \ch{Cl} cages surrounding them is preserved during the distortion. The odd loops, however, introduce frustration in the structure and the octahedral cages cannot always share one edge. This can be appreciated in Fig.~\ref{fig:DFT}~(b), where the two octahedral cages in the distorted bond get separated. The determination of the extent to which this frustration is detrimental to the directional Kitaev exchange goes beyond the scope of this work.     

\begin{figure}[!h]
\includegraphics{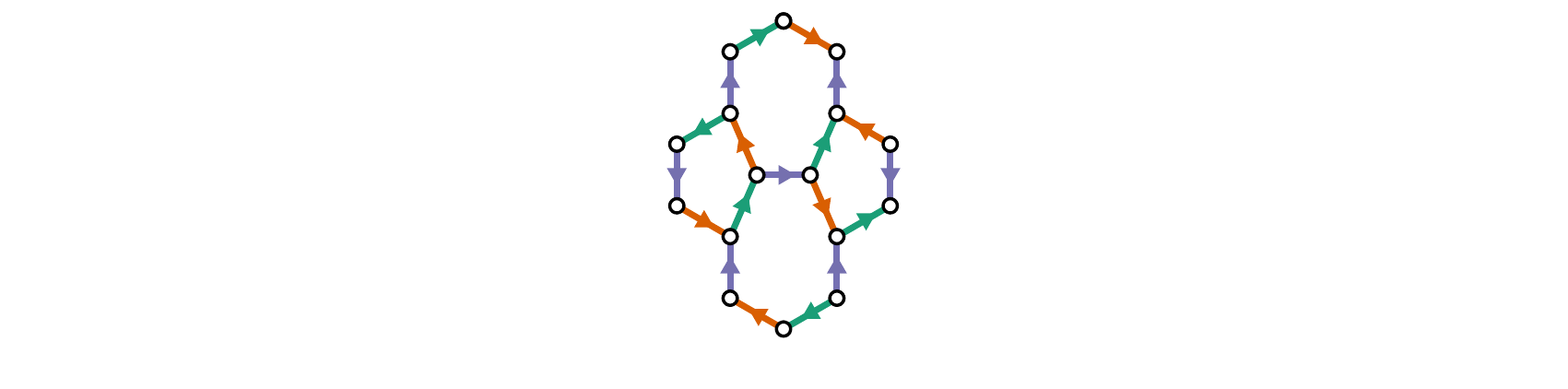}
\caption{
Unit cell with arrows indicating the positive orientation of the bilinears $u_{jk}$. Bond colors highlight the type of spin-spin coupling across a bond $\sigma^{\alpha}_{j}\sigma^{\alpha}_{k}$, $\alpha = x, y, z$ (violet for $z$, orange for $y$ and green for $x$). } 
\label{fig: bond direction}
\end{figure}

\vskip 15pt
\begin{center}
\textit{Convention for bond orientation}
\end{center}
Majorana operators satisfy the anticommutation relations $\{ b_i^{\alpha},b_j ^{\alpha}\}=0$ for $i \neq j$. Thus, the Majorana bilinears $u_{jk}=\mathrm{i}b_j^{\alpha_{jk}}b_k^{\alpha_{jk}}$ are odd under index permutation: $u_{jk}=-u_{kj}$. The order in the product of Eq.(4) of Main Text forces us to choose an arbitrary convention for the orientation of the bonds. The convention adopted throughout this work is presented in Fig.~\ref{fig: bond direction}, where the $u_{jk}$ are taken positive in the arrow's direction. Note how, in accordance with Eq.(2) of Main Text, each loop contains an even number of bonds directed clockwise. With this convention, the plaquette operators defined in Eq.~(2) of Main Text ensure that the composition law for fluxes holds~\cite{PhysRevB.90.134404}: $W_{1+2}=W_1W_2$.

\vskip 15pt 
\begin{center}
\textit{Effects of the couplings sign}
\end{center}
As argued in the Main Text, the sign of the couplings has no effects on the energy of the system. A change in the sign of $J_x$ or $J_y$ can be reabsorbed changing the sign of an even number of $u_{jk}$, leaving the eigenvalues of $W_p$ unchanged and without exiting the flux sector. At the same time, the model with a negative $J_z$ can be mapped to a configuration with a positive coupling and an odd number of $u_{jk}$’s with flipped signs per plaquette. This maps each flux sector to its time reversal partner and does not change the energy.

\begin{figure}[!h]
\includegraphics{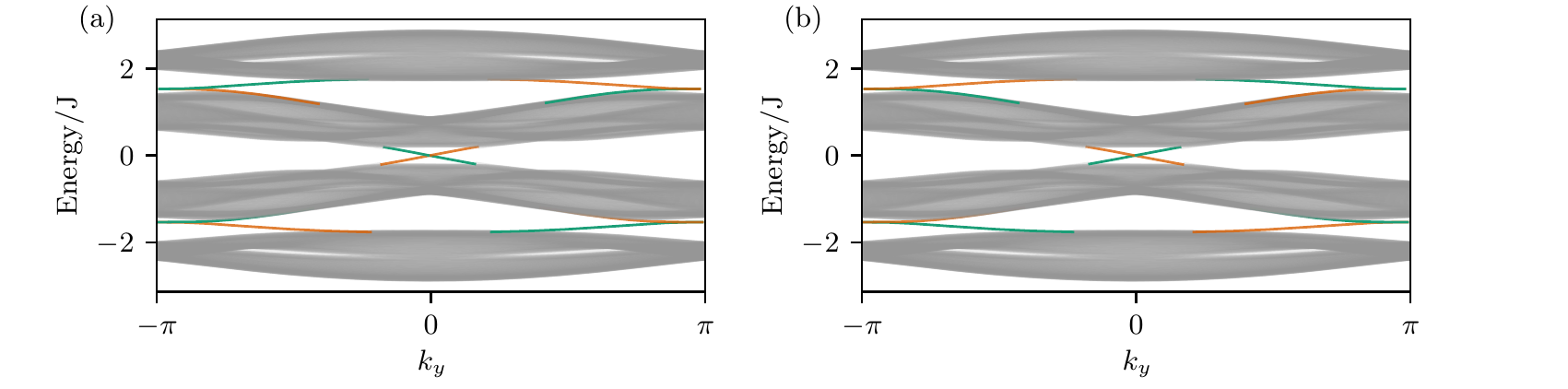}
\caption{
Boundary spectrum of the flux free sector for a ribbon with open boundary
conditions along $\bs{e}_1$ and periodic along $\bs{e}_2$, with $J_x = J_y = |J_z|
= J$. The ribbon is composed of 50 unit cells along $\bs{e}_1$. In orange the modes located on the edge with normal $+\bs{e}_1$ and in green edge modes located on the opposite edge. (a) $J_z>0$ (as in the Main Text), the Chern
numbers of the three gaps are  $C=-1,+1,-1$ from bottom to top. (b) $J_z<0$, the Chern
numbers of the three gaps are  $C=+1,-1,+1$ from bottom to top. In the two cases, edge modes on the same surface propagate in opposite directions. 
}  
\label{fig: coupling sign}
\end{figure}

A change of sign can however affect other quantities. For example, changing the sign of $J_z$ changes the sign of the Chern number in each gap of the spectrum. In fact, the time reversal operation that absorbs the change $J_z\rightarrow -J_z$ flips the sign of the Chern number. In Fig.~\ref{fig: coupling sign}, we show how the same system with different signs of $J_z$ has edge modes propagating in opposite directions.

\begin{figure}[!h]
 \includegraphics{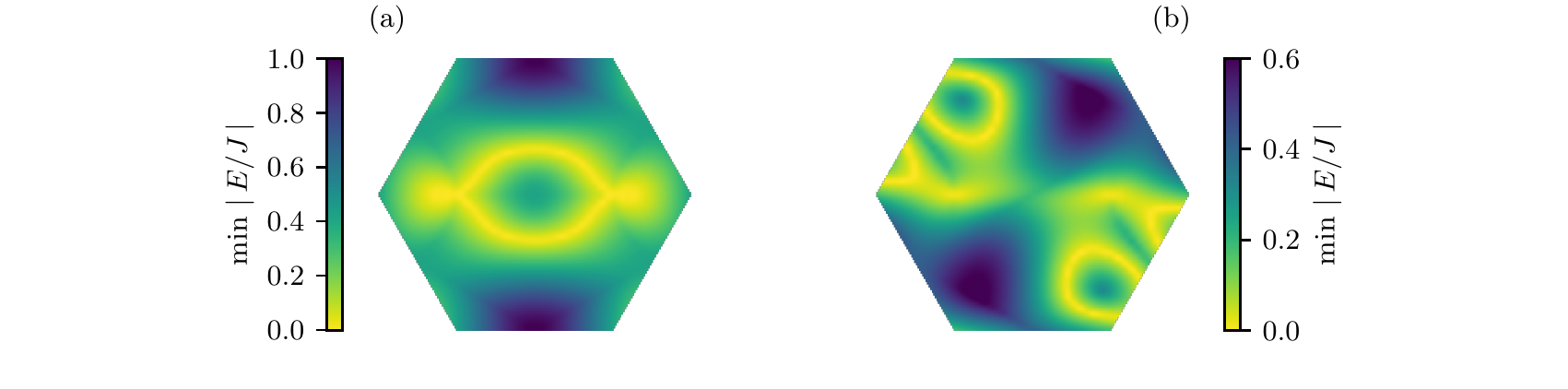}
 \caption{
Fermi surface in the first Brillouin zone. The plot shows the minimum of the absolute value of the eigenvalues in the first Brillouin zone of the pentaheptite lattice and the line of zero energy defines the Fermi surface. (a) Free-Fermions. (b) Free-Majoranas in the presence of $\pi$ flux in a pentagon and a heptagon of the unit cell. In both cases $J_x=J_y=J_z=J$.
 }  
 \label{fig:MajoFermi}
 \end{figure}

\vskip 15pt
\begin{center}
\textit{Majorana Fermi surface}
\end{center}
The free-fermions tight-binding model on a pentaheptite lattice has a metallic character. The Fermi surface at half filling is shown in Fig.~\ref{fig:MajoFermi}~(a).
On the other hand, in the zero flux sector of the Kitaev model, the single particle Majorana spectrum is gapped. In this section, we do not deal with the full many-body problem. Rather we study the band dispersion of free Majorana fermions moving in a static background field and relax the constraint of conserved fermionic parity. One may wonder whether there are simple configurations of static background $\mathbb{Z}_2$ fields that close the gap and induce a Majorana Fermi surface. We answer such curiosity in the affirmative. In particular, the flux configuration that breaks $C_2$ symmetry and preserves the full lattice translation symmetry, without enlarging the unit cell, achieves this goal. With a $\pi$ flux in a heptagon and a pentagon of the unit cell, the spectrum develops a Majorana Fermi surface, as shown in Fig.~\ref{fig:MajoFermi}~(b). 
 
 \vskip 15pt
 \begin{center}
 \textit{Perturbation theory in the $A_1$ and $A_2$ phases}
  \end{center}
 Phases $A_1$ and $A_2$ are conveniently understood in a limit where
 one of the couplings $J_x$, $J_y$, or $J_z$ is much larger than the
 others. This is a good starting point for a perturbation theory in the Majorana fermion representation. Compared to the Rayleigh-Schr\"odinger perturbation theory for spins, the same analysis in the fermionic representation can be formulated in the Feynman diagrams language~\cite{PhysRevB.90.134404}. This results in few simple diagrammatic rules. For sake of concreteness, we consider the case $J_z\gg J_x=J_y$ and present the rules in this limit. (i) Construct all possible closed paths involving weak and strong bonds. (ii) Compute the amplitude of the path. Each weak bond $ij$ contributes a factor $2J_xu_{ij}$. Each strong bond contributes a factor $2J_zu_{ij}/\left(\omega^2+(2J_z)^2\right)$. Each strong bond attached to the path with only one site gives a factor $\omega/\left(\omega^2+(2J_z)^2\right)$. Give an extra factor $1/2$ and integrate over the whole frequency range $-\infty \leq \omega \leq +\infty$. (iii) Sum over all possible paths considering the reverse of a non-contractible path as a different one. 

Few observations can be readily drawn from these rules. Self-retracting paths gives energy contributions that are flux independent. In fact, each bond $ij$ appears twice giving a constant contribution: $u_{ij}u_{ji}=-1$. Non-trivial contributions are given by non-contractible loops. Among those, only the even length ones contribute. In fact, since $u_{ij}=-u_{ji}$, loops of odd length followed in opposite directions result in opposite sign contributions that cancel each other. A closed loop with $n$ weak bonds will give a contribution of order $J_x^n$. The amplitude of a path of length $\ell$ is a positive number times $(\mathrm{+i})^\ell W_\ell$. We now focus our attention on non-self-retracting loops that give the first non-trivial corrections in the perturbative Hamiltonians for the $A_1$ and $A_2$ phases of the Kitaev model on the pentaheptite lattice.

\begin{figure}[!h]
\includegraphics{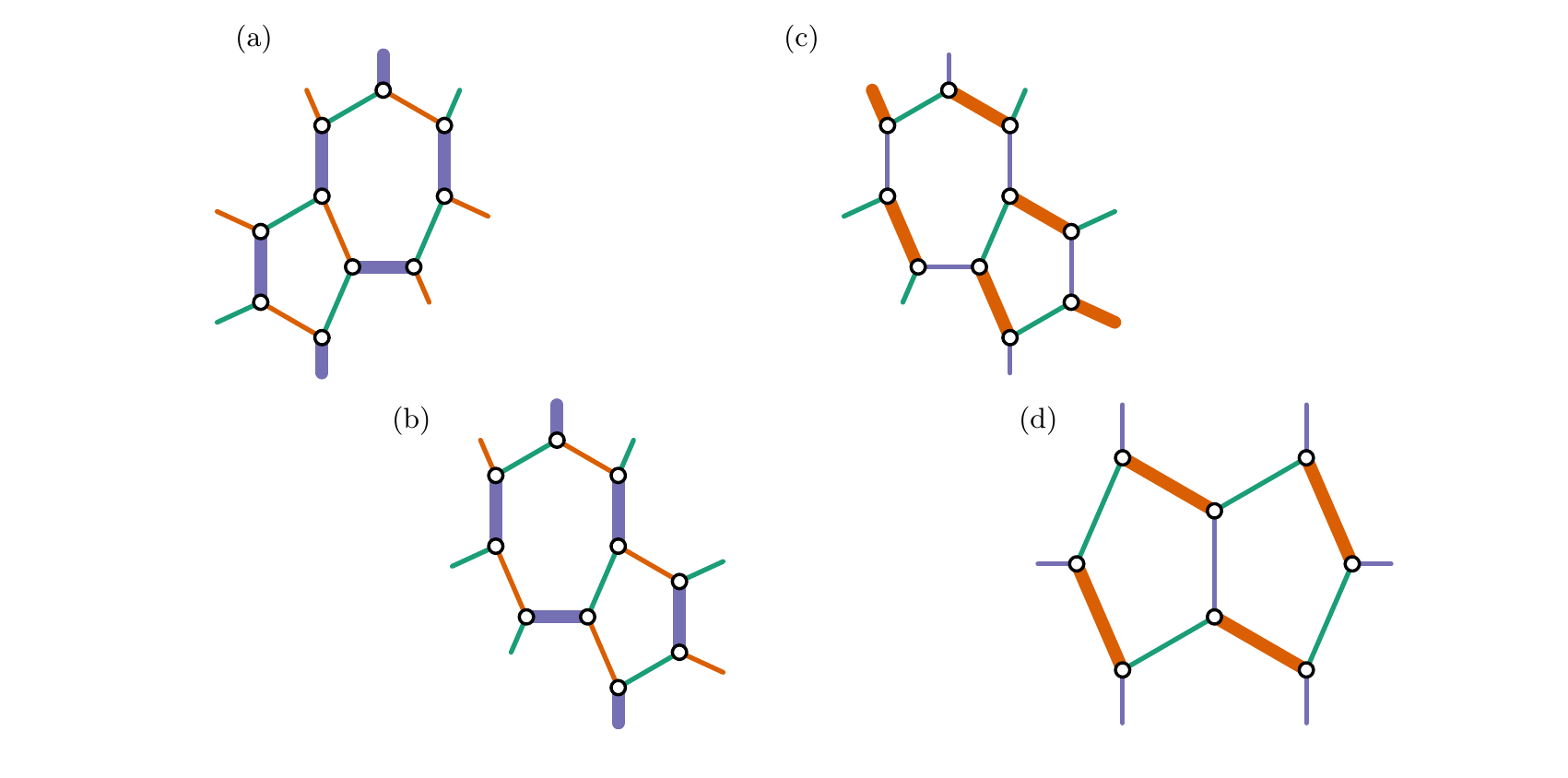}
\caption{
Non-self-retracting paths on the pentaheptite lattice. The thickness of the bond is proportional to the coupling strength. Bond colors highlight the type of spin-spin coupling across a bond $\sigma^{\alpha}_j\sigma^{\alpha}_k$, $\alpha = x, y, z$ (violet for $z$, orange for $y$ and green for $x$). (a)--(b) Inequivalent loops of length 10 that give the sixth order perturbation Hamiltonian for the $A_1$ phase. (c) Loop of length 10 that gives the sixth order contribution to the perturbative Hamiltonian in the phase $A_2$. (d) Loop of length 8 that gives the flux-dependent fourth order contribution. 
}  
\label{fig:perturbation}
\end{figure}

For the $A_1$ phase, we consider the limit $J_z\gg J_x,J_y$ and all positive couplings. The lowest order flux-dependent contribution is given by the loops of length 10 enclosing a pentagon and a heptagon. There are two inequivalent loops of this type per unit cell, each with double multiplicity (see Fig.~\ref{fig:perturbation}). The contribution from the loop of Fig.~\ref{fig:perturbation}~(a) is:
\be
-2\times 2
\label{HeptCost} \int_{-\infty}^{\infty}\frac{\text{d}\omega}{2\pi}\frac{1}{2}(2J_x)^4(2J_y)^2\frac{\omega}{\omega^2+(2J_z)^2}\frac{\omega}{\omega^2+(2J_z)^2}\frac{(2J_z)^4}{\left(\omega^2+(2J_z)^2\right)^4}W_5W_7=-\frac{7J_x^4J_y^2}{128J_z^5}W_5W_7\,,
\ee
where a factor $2$ comes from the same loop traversed in opposite direction and another factor $2$ from the presence of two different loops of this type per unit cell. Here we use $W_{10}=W_5W_7$, according to our definition of the plaquette operators (see Main Text). The contribution from the similar loop of Fig.~\ref{fig:perturbation}~(b) reads:
\be
-2\times 2 \int_{-\infty}^{\infty}\frac{\text{d}\omega}{2\pi}\frac{1}{2}(2J_x)^2(2J_y)^4\frac{\omega}{\omega^2+(2J_z)^2}\frac{\omega}{\omega^2+(2J_z)^2}\frac{(2J_z)^4}{\left(\omega^2+(2J_z)^2\right)^4}W_5W_7=-\frac{7J_x^2J_y^4}{128J_z^5}W_5W_7\,.
\ee
Summing all these terms with the correct multiplicity, we get the sixth order effective Hamiltonian for the $A_1$ phase:
\be
H^{(6)}_{\text{eff}}=\text{const.}\,-\frac{7}{128}\left(\frac{J_x^4J_y^2}{J_z^5}+\frac{J_x^2J_y^4}{J_z^5}\right)W_5W_7\,.
\ee
 
 A similar analysis can be performed for the phase $A_2$ in the limit $J_y\gg J_x,J_z$. The lowest order non-contractible loop is of length 8 and encloses two pentagons, Fig.~\ref{fig:perturbation}~(d). It gives a fourth order contribution:
\be
\label{pentCost}
2\int_{-\infty}^{\infty}\frac{\text{d}\omega}{2\pi}\frac{1}{2}(2J_x)^4\frac{(2J_y)^4}{\left(\omega^2+(2J_y)^2\right)^4}W_5W_{5'}=\frac{5J_x^4}{16J_y^3}W_5W_{5'}\,,
\ee 
where the factor 2 comes from the two orientations of the loop and we used $W_8=W_{5}W_{5'}$. Eq.~\eqref{pentCost} gives the dominant energy cost for an isolated vortex in a pentagonal plaquette. It does not, however, determine the energy cost of a vortex in a heptagonal plaquette. Hence, we need to consider sixth order contributions given by loops of length 10 enclosing a pentagon and a heptagon (see Fig.~\ref{fig:perturbation}~(c)). The contribution of such loops is given by:
\be
-2\times 2 \int_{-\infty}^{\infty}\frac{\text{d}\omega}{2\pi}\frac{1}{2}(2J_x)^2(2J_z)^4\frac{\omega}{\omega^2+(2J_y)^2}\frac{\omega}{\omega^2+(2J_y)^2}\frac{(2J_y)^4}{\left(\omega^2+(2J_y)^2\right)^4}W_5W_7=-\frac{7J_x^2J_y^4}{128J_z^5}W_5W_7\,,
\ee
similarly to Eq.~\eqref{HeptCost}. Summing all contributions gives the sixth order perturbative Hamiltonian for the $A_2$ phase:
\be
H^{(6)}_{\text{eff}}=\text{const.}\,+\frac{5J_x^4}{16J_y^3}W_5W_{5'}-\frac{7J_x^2J_z^4}{128J_y^5}W_5W_7\,.
\ee

Note that in the vortex-free sector all heptagonal plaquettes have eigenvalue $+\mathrm{i}$ and the pentagonal ones $-\mathrm{i}$. Hence, in sixth order perturbation theory, isolated vortices have a finite energy cost over the ground state energy in phase $A_1$ and $A_2$.

\vskip 15pt
\begin{center}
\textit{Phase transition between $A$ and $B$ phases}
\end{center}
The Kitaev model on the pentaheptite lattice realizes two different types of spin liquids. In the phase $B$, one finds a non-Abelian chiral spin liquid. In phases $A_1$ and $A_2$, the ground state of the model is an Abelian spin liquid. These phases are separated by the phase boundaries defined in Eq.(5) of the main text. There, the gap in the single particle Majorana spectrum closes, while the flux-gap remains open. 

\begin{figure}[!h]
\includegraphics{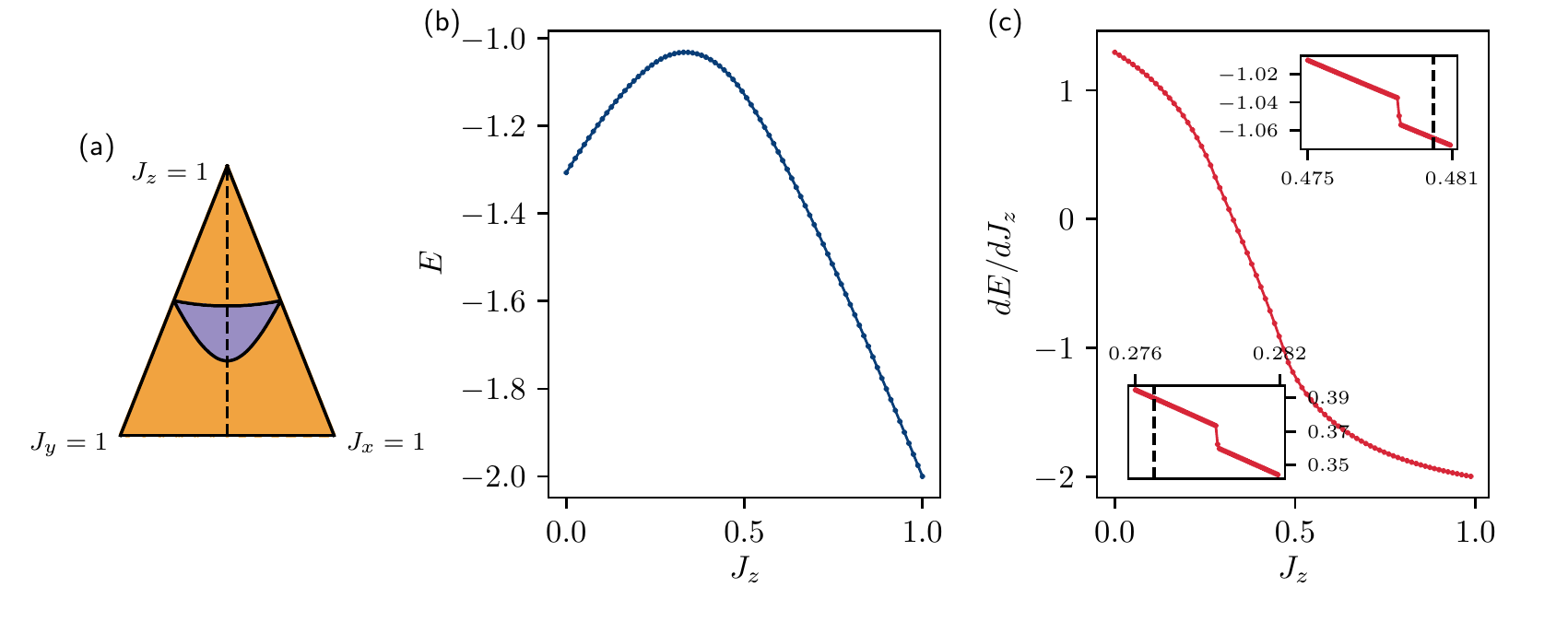}
\caption{
Phase transitions between $A$ and $B$ phases. (a) Phase diagram as in Fig.2(a) of the main text. In violet the region of phase $B$ and in orange of phases $A_1$ and $A_2$. The black dashed line indicates the coupling values considered to study the phase transitions. (b) Ground state energy for a sample on a torus with $L=10$ as a function of $J_z$. The other coupling constants are $J_x=J_y=(1-J_z)/2$. (c) First derivative of the ground state energy with respect to the coupling $J_z$. The lower inset shows a zoom-in around the transition between $A_2$ and $B$. The theoretical value $J_z=0.277$ at which the phase transition should occur is highlighted by a black dashed line. Finite size effects shift the discontinuity of the first derivative associated to the first-order phase transition. The upper inset shows a similar zoom-in for the $A_1$--$B$ transition, predicted to occur at $J_z=0.480$.  
}  
\label{fig:Transition}
\end{figure}

Tuning $J_z$ as shown in Fig.~\ref{fig:Transition}(a), one can explore the whole phase diagram and compute the energy of the flux-free sector. We perform such analysis for a finite system on a torus with $L=10$, cf. Fig.~\ref{fig:Transition}(b). As it can be seen in  Fig.~\ref{fig:Transition}(c), the derivative of the ground state energy with respect to $J_z$ shows a discontinuity when crossing from the $A_2$ phase to the $B$ one and from $B$ to $A_1$. The location of the jump in the first derivative is affected by finite size effects but occurs at $J_z$ values close to the theoretically predicted $J_z=0.277$ and $J_z=0.480$. This observation proves that the phase transitions at the phase boundaries of Eq.5 of the main text are topological first-order phase transitions.

\vskip 15pt
\begin{center}
\textit{Topological phases degeneracy}
\end{center}
We checked the degeneracies of the different topological phases in the vortex-free sector. Only physical state with fixed fermionic parity have been considered. In the $A_1$ and $A_2$ phases, we find a four-fold degeneracy of the ground state on a torus. This is consistent with the Abelian topological phase of $\mathbb{Z}_2$ topological order. The four different ground states (GS) are fully characterized by the $\mathbb{Z}_2$ global fluxes $(\phi_x\,,\phi_y)$ around the handles of the torus. In the non-Abelian phase $B$, the GS degeneracy is threefold, compatible with an Ising topological field theory. One of the flux-free sectors of the non-Abelian phase does not belong to the physical subspace (with our convention, the state $(0,0)$ does not belong to the physical subspace). 

It is important to stress how the exact GS degeneracy is present only in the thermodynamic limit. For example, on the torus $L\bs{e}_1\times L\bs{e}_2$ the three GS of the non-Abelian phase have considerably different energies for $L=2$. For $J_x=J_y=J_z=J$, the energy per unit cell in units of $J$ is:
\begin{center}
\begin{itemize}
\item (1,0): -3.1044
\item (0,1): -3.1044
\item (1,1): -3.0892
\end{itemize}
\end{center}
The first excited state has lower energy than the vortex-free configuration $(1,1)$. However, for $L=10$ the system is already large enough to show a convergence of the energies of the voretx-free states:
\begin{center}
\begin{itemize}
\item (1,0): -3.097073
\item (0,1): -3.097073
\item (1,1): -3.097072
\end{itemize}
\end{center}

\begin{figure}[!h]
\includegraphics{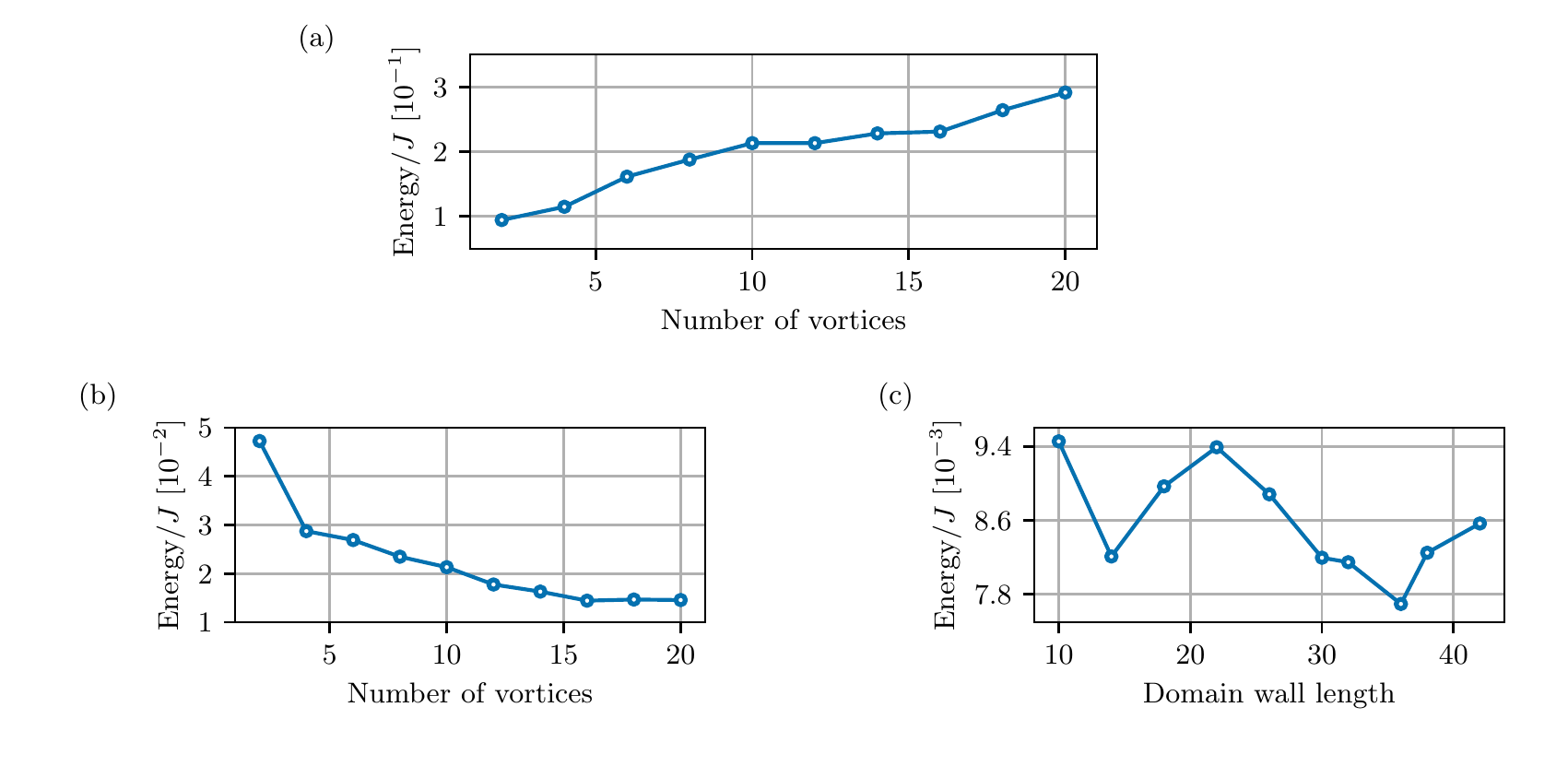}
\caption{
Energy cost for vortex clusters of different sizes. (a) Global energy cost over the vortex-free flux sector as a function of the number of vortices in the cluster. (b) Energy cost per single vortex as a function of the number of vortices in the cluster. (c) Energy cost for unit length of cluster's domain wall.
}  
\label{fig: Vortex cluster}
\end{figure}

\vskip 15pt
\begin{center}
\textit{Vortex energies}
\end{center}
We studied in further details the vortex sector in the $B$ phase. The nucleation of a well isolated vortex costs a non-zero amount of energy. However, a pair of vortices close to each others has little energy cost and it suggests the existence of an attractive force between vortices that could favor the formation of clusters. We study the energy cost of clusters on a torus $L\bs{e}_1\times L\bs{e}_2$ with $L=10$ and $J_x=J_y=J_z=J$. In Fig.~\ref{fig: Vortex cluster}~(b) we see how the energy cost per vortex decreases while the cluster size increases. This seems to point to the existence of a GS different from the vortex-free one. However, Fig.~\ref{fig: Vortex cluster}~(a) shows the finite energy gap between the vortex-free sector and the configuration with a vortex cluster. The decrease in energy cost per single vortex is not enough to compensate the increasing number of vortices in the cluster. Therefore, the formation of large clusters is not favored. As a last evidence, in Fig.~\ref{fig: Vortex cluster}~(c), we calculate the cost per unit length of the domain wall between the cluster and the rest of the system. This quantity fluctuates around a constant non-zero value and suggests the existence of a finite energy cost to create a domain wall. All these results seem to validate the assumption of the vortex-free flux sector as ground state. 

Finally, we checked wether ``stripy-like'' states as the first excited states of the sample with $L=2$ (cf. Fig.~2(c) of main text) are favorable in larger samples. When $L=10$ the energy cost of two vortices in neighboring pentagonal plaquettes is $0.0852$. A single stripe crossing the sample as in Fig.~2(c) of main text has a cost of $0.4728$. We further checked that the energy cost of a larger stipe in the middle of the sample that creates excitation in half of the sample's plaquettes is $0.5779$. Finally a configuration with alternating single stripes excitation costs $1.9993$. These observations suggest that in larger samples the first excited state is realized a by a pair of neighboring fluxes. The case of a smaller sample $L=2$ is with this regard special as it is mainly affected by finite size effects. These are particularly dramatic for $J_z>J_x\,,J_y$. When this condition is met, the ground state of the system with $L=2$ is in the flux configuration of Fig.~2(c) of the main text rather than in the vortex-free one. We carefully checked that such a situation is realized only in the small sample with $L=2$ and disappears already for $L=3$. Therefore, we stress that finite size effects are particularly relevant for a small sample $L=2$ and should be treated carefully.

Our findings call for potential extensions of exact results for the Kitaev models which are based on reflection positivity, which is not fulfilled by the pentaheptite lattice.

\begin{figure}[!h]
\includegraphics{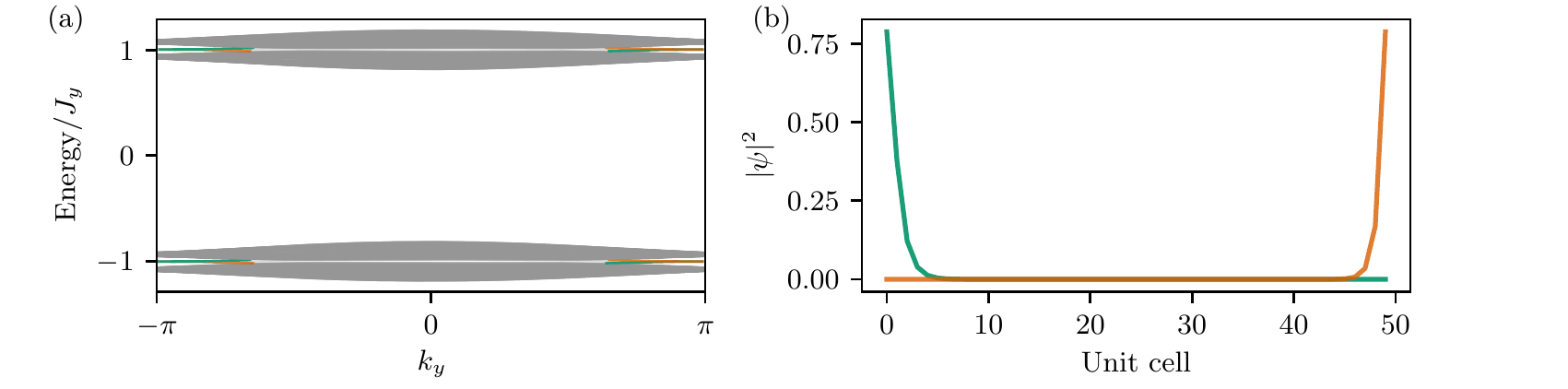}
\caption{
(a) Boundary spectrum of the vortex-free sector in the Abelian phase $A_2$ for a ribbon with open boundary
conditions along $\bs{e}_1$ and periodic along $\bs{e}_2$, with $J_y=1$, $J_z=J_x=0.1$. The ribbon is composed of 50 unit cells along $\bs{e}_1$. There are no edge modes crossing the zero energy single-particle gap. (b) Square modulus of the wavefunctions for the edge modes in Fig.~3~(a) of Main Text. The green state is strongly localized on the edge with normal $-\bs{e}_1$, while the orange mode is localized on the opposite edge.
}  
\label{fig: Abelian modes}
\end{figure}

\vskip 15pt
\begin{center}
\textit{Edge modes}
\end{center}
In Fig.~\ref{fig: Abelian modes}~(a) we show the spectrum of a ribbon for the Abelian phase $A_2$. There are no chiral Majorana modes crossing the band gap around zero energy.

The non-Abelian phase $B$ is characterized by $|C|=1$ and a chiral Majorana edge mode crosses the zero energy gap, as shown in Fig.~\ref{fig: coupling sign}. In Fig.~\ref{fig: Abelian modes}~(b), we show the localization of these modes at the boundary of the ribbon.

	%\putbib[biblioNew]

	\end{bibunit}

	\end{document}